# Low Threshold Two-Dimensional Annular Bragg Lasers

Jacob Scheuer[1], William M. J. Green[2], Guy DeRose[2], Amnon Yariv[1,2]

[1]Department of Applied Physics, 128-95 California Institute of Technology,

Pasadena, California 91125

[2]Department of Electrical Engineering, 128-95 California Institute of Technology,

Pasadena, California 91125

koby@caltech.edu

**Abstract**

Lasing at telecommunication wavelengths from annular resonators employing radial Bragg reflectors is demonstrated at room temperature under pulsed optical pumping. Sub milliwatt pump threshold levels are observed for resonators with 0.5-1.5 wavelengths wide defects of radii 7-8 $\mu$m. The quality factors of the resonator modal fields are estimated to be on the order of a few thousands. The electromagnetic field is shown to be guided by the defect. Good agreement is found between the measured and calculated spectrum.

Ring resonators are versatile elements with various applications ranging from telecommunication and sensing to basic scientific research [1-5]. During the last few years, considerable effort has been focused on improving the quality factors (Q) of resonators and reducing their modal volume (see e.g. Refs. [5, 6] and references therein).

Recently, we proposed a novel type of annular resonator based on radial Bragg reflection [7, 8]. These devices, named annular Bragg resonators (ABR), offer smaller dimensions compared to conventional resonators which employ total internal reflection (TIR), while retaining high Q. In addition, such structures exhibit superior sensitivity compared to conventional resonators for biological and chemical sensing applications [9]. This class of resonators is closely related to the family of circular-gratings DBR resonators which also exhibit lasing patterns with low angular propagation coefficients [10, 11].

In this letter, we report on the observation of photoluminescence and lasing from ABRs realized in semiconductor material (see Fig. 1). The structure consisted of a circumferentially guiding defect surrounded by radial Bragg reflectors. Due to the circular geometry, the optimal layer widths required to confine the light in the defect are not constant but rather monotonically decreasing with the radial distance. The widths of the layers are determined by the zeros and extrema of the Bessel function of order $m$ where $m$ is the angular propagation coefficient of the mode for which the device was designed. For simplicity, we label the separation between a zero and the nearest extremum and between successive zeros as "quarter-wavelength" and "half-wavelength" respectively. Although these distances are not constant across the device, their role in the construction of the distributed reflector is similar to that of their Cartesian counterparts [12]. The reflector layers could also be of higher Bragg order (i.e., $3/4\lambda$, $5/4\lambda$, etc. in the same notation). Although such an approach requires more Bragg layers to confine the light in the defect (compared to reflector based on $\lambda/4$ layers), it facilitates the fabrication of the devices. The defect width of an ABR is a multiple integer of "half-wavelength", meaning its interfaces are located at zeroes of the field.

ABRs of several geometries and Bragg reflector orders were fabricated within a 250nm thick membrane of InGaAsP with six 75Å quantum wells (QW) positioned at the center. After the ABR patterns were etched into the active material, the original

InP substrate was removed and the membrane was transferred to a sapphire plate using an ultraviolet curable optical adhesive to improve the vertical confinement of the electromagnetic field in the device [13].

To simplify the design and the modeling of the devices we employed the effective index approximation in the vertical dimension. The effective index of the membrane was found to be approximately 2.8 for the $H_z$ polarization and 2.09 for the $E_z$ polarization. Since the $H_z$ polarization is more confined than the $E_z$ polarization and the optical gain of the compressively strained QW structure used favors the $H_z$ polarization [14], we optimized the radial structure to this polarization [8].

To facilitate the fabrication of the devices, a mixed Bragg order approach was adopted [8]. The high-index Bragg layers ($n_{eff}$=2.8) were 3/4λ wide (second order) and the low-index layers ($n_{eff}$=1.0 for air gaps and $n_{eff}$=1.54 for adhesive filled gaps) were λ/4 wide (first order). In addition to the relaxed fabrication tolerances, such a layer structure improves the vertical confinement (larger material filling factor) and induces efficient vertical emission. While the latter decreases the overall Q-factor of the cavity, it also enables simple observation of the intensity pattern which evolves in the device [15]. Figure 1 depicts SEM micrographs of some of the fabricated devices. The specific parameters of each structure are detailed in the figure caption.

The near-field (NF) intensity pattern and the emitted spectrum of the ABRs were examined at room temperature under pulsed optical pumping. The pump beam was focused through the transparent sapphire substrate on the backside of the sample. Half of the pump beam intensity was split by a 3dB beam-splitter and was focused on a detector to obtain the pump power. The vertical emission from the front side of the sample was wither focused on an IR camera to obtain the NF intensity pattern or coupled into a multi-mode fiber to obtain the spectrum.

The spectrum emitted from the pumped, unpatterned, QW layer structure consisted of a wide peak centered at 1559nm. When an ABR was pumped, the emission characteristics changed significantly. While the specific details varied from device to device, the overall behavior was similar. Once a certain pump intensity threshold was exceeded, clear and narrow emission lines appeared in the spectrum (see Fig. 2). As the pump intensity was increased, the intensity of the emission lines increased while broadening towards shorter wavelengths. Increasing the pump power further resulted in the appearance of additional emission lines.

Figure 2 depicts the emitted spectra from an ABR for various pumping levels. The specific device consisted of a half-wavelength wide defect surrounded by five and ten reflection gratings periods in the inner and outer sides respectively. The inset of Fig. 2 shows an L-L curve of the same device, indicating a threshold at $P_{th}$=1.0mW. Other devices exhibited even lower threshold levels, the lowest being 0.6mW.

At threshold, emission lines spaced approximately 14nm apart appeared in the spectrum. At twice the threshold level, two additional emission lines appeared at $\lambda$=1.608µm and $\lambda$=1.623µm. At $P_{pump}$=2.4mW three additional emission lines appeared at $\lambda$=1.593µm, $\lambda$=1.612µm and $\lambda$=1.626µm. Increasing the pump even further resulted only in variation of relative intensities of the emitted modes.

In order to understand the spectral characteristics, a finite difference time domain (FDTD) simulation tool was used to model the device [16]. Figure 3 shows a comparison between the measured (A) and the calculated (B) spectra. Good agreement was found between the measured and calculated spectra not only for the resonance wavelength but also for the relative amplitudes. It should be noted that the measured spectrum is to some extent "compressed" compared to the calculated one. This is because the FDTD model does not account for the material dispersion of the membrane. The field profiles of the various modes, can be classified into three distinct categories (see Fig. 3A). The modes belonging to class D are confined within the defect with different angular propagation coefficients. The modes belonging to class I are confined in the ring closest to the defect from the internal side. These modes are actually guided by TIR and are supported by the structure because of the use of second order high-index layers. The rest of the modes, labeled by M are not localized in a specific layer but are distributed over several grating periods, peaking both in the defect and in one of the rings of the internal reflector. Unlike the I class, the M family modes are confined by Bragg reflection. The angular propagation factors of $M_1$ and $M_2$ are 41 and 31 respectively, corresponding to effective indices of 1.48 and 1.12, defined by $n_{eff} = m\lambda/2\pi\rho_{eff}$ where $\lambda$ and $\rho_{eff}$ are respectively the resonance wavelength and the effective radius of the modes. These effective indices are lower than the refractive index of the adhesive filling the gaps, which confirms that the radial confinement mechanism is indeed Bragg reflection.

Figure 4 shows the intensity pattern emitted from the device at a pump level of 1.6mW. The emitted pattern consisted of two concentric and relatively wide rings of light. The radius of the outer ring corresponds to the radius of the defect, indicating that the lasing pattern belongs to the M family.

In summary, we demonstrated annular Bragg lasers realized in a thin membrane of InGaAsP active semiconductor material. Lasing was achieved at room temperature under pulsed optical pumping conditions. Sub mW threshold levels were observed for ABRs with 7-8μm defect radii pumped by a 30μm diameter spot. Imaging the vertical IR emission from the devices indicated localization of the field within the defect.

The authors would like to thank Dr. Axel Scherer and Dr. Oskar Painter for fabrication facilities access and George Paloczi and Dr. Reginald Lee for fruitful discussions. This work was supported by the NSF, DARPA and AFOSR.

# Figure Caption

Figure 1 – Scanning electron microscope images of the fabricated ABRs; A) $\lambda/2$ wide high-index defect. The defect is the slightly narrower $6^{th}$ ring from the center, marked by the arrow; B) $3\lambda/2$ wide high-index defect; C) $3\lambda/2$ wide air defect; D) Close-up of the reflector layers.

Figure 2 – The evolution of the emitted spectrum from the ABR shown in Fig. 1(A) as a function of the pump intensity. Inset – L-L curve of the device.

Figure 3 – Measured (A) and calculated (B) spectral response of the ABR shown in Fig. 1(A). D modes are confined in the defect; I modes are confined in one of the rings of the internal reflector; M modes are not confined to a single ring.

Figure 4 – IR image of the vertical emission from an ABR. The lasing pattern corresponds to mode $M_2$.

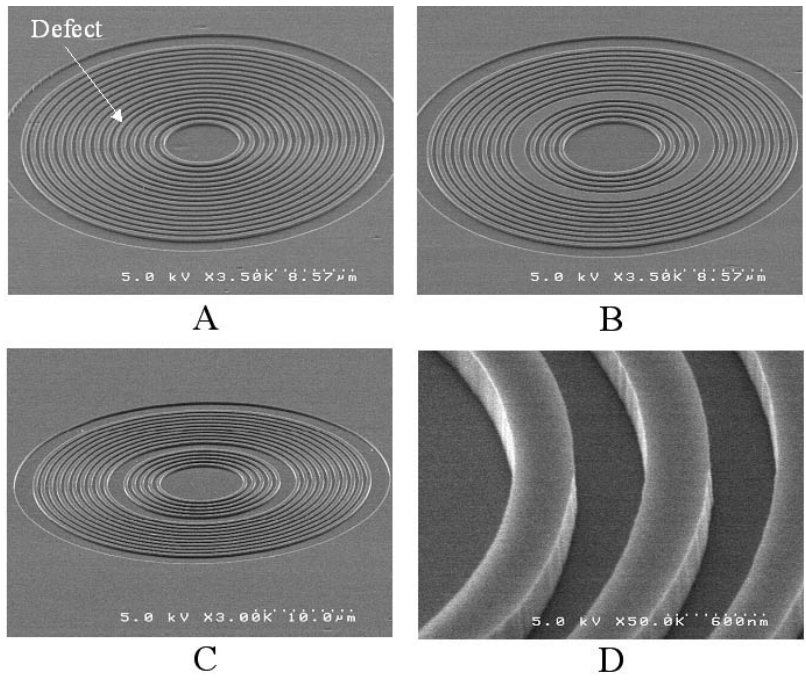

Figure 1

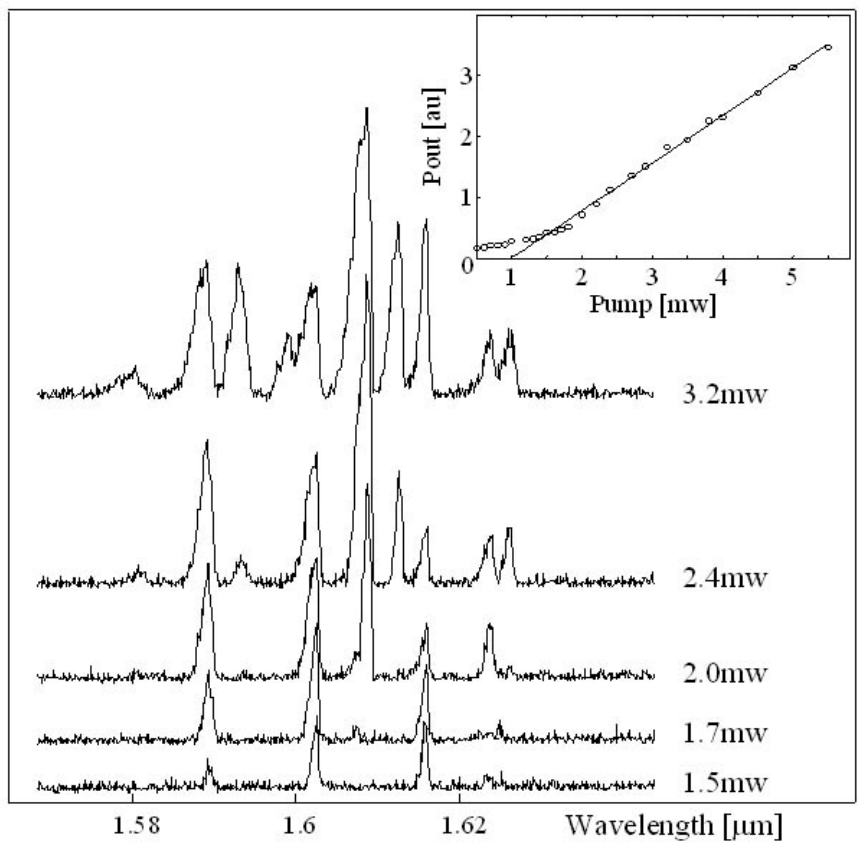

Figure 2

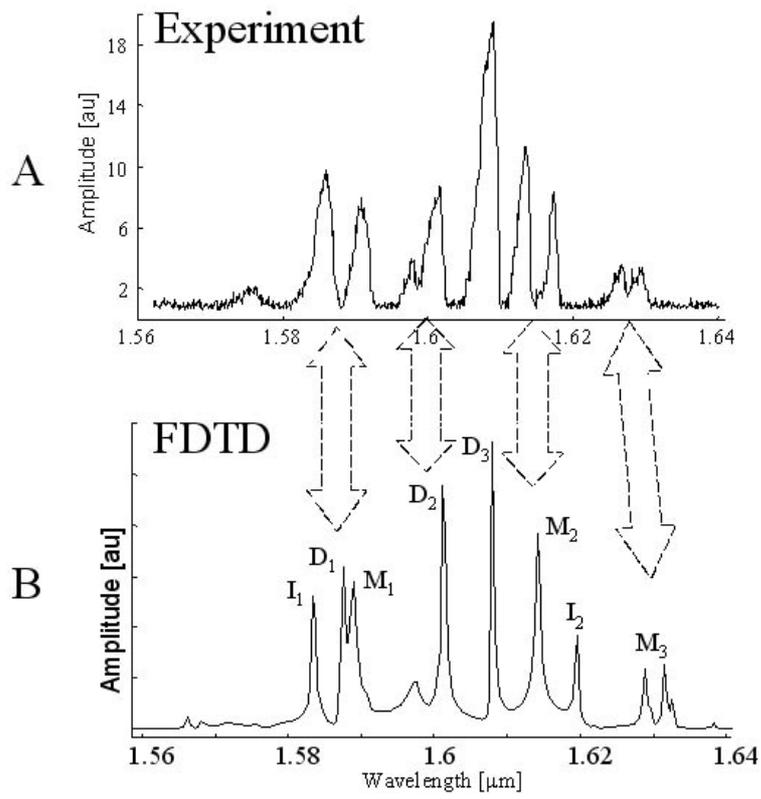

Figure 3

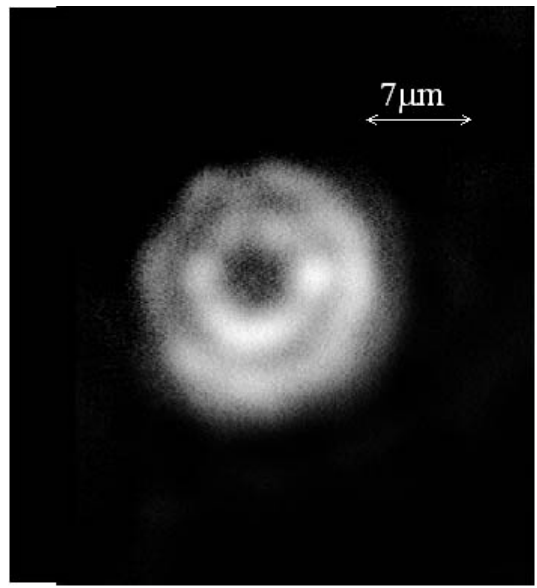

Figure 4